\documentstyle[prl,twocolumn,aps,psfig]{revtex}
\begin{document}
\draft
\twocolumn[\hsize\textwidth\columnwidth\hsize\csname
@twocolumnfalse\endcsname
\title{Investigation of Single Boron Acceptors at the Cleaved Si:B (111) 
Surface}
\author{Maya Sch\"ock, Christoph S\"urgers, Hilbert v. L\"ohneysen}
\address{Physikalisches Institut, Universit\"at Karlsruhe, D-76128 Karlsruhe,
Germany}
\maketitle
\begin{abstract}

The cleaved and ($2 \times 1$) reconstructed (111) surface of $p$-type Si is investigated
by scanning tunneling microscopy (STM). Single B acceptors are identified due to 
their characteristic voltage-dependent contrast which is explained by a local 
energetic shift of the electronic density of states caused by the Coulomb 
potential of the negatively charged acceptor. In addition, detailed analysis
of the STM images shows that apparently one
orbital is missing at the B site at sample voltages of $0.4 - 0.6$ V, corresponding to the absence of a localized
dangling-bond state. Scanning tunneling spectroscopy (STS) confirms a strongly altered
density of states at the
B atom due to the different electronic structure of B compared to Si.

\end{abstract}
]

\section{Introduction}
In recent years much interest 
has been devoted to identify individual dopant atoms at the surface of doped 
semiconductors by scanning tunneling microscopy (STM). 
For instance, on GaAs (110) 
substitutional Si donors (Si$_{\rm Ga}$) and Be or Zn acceptors in the top 
few surface layers appear as protrusions due to the local change of the 
tip-induced band-bending arising from the Coulomb
potential of the ionized dopant \cite{Zheng94,Johnson93}.
At low temperatures and negative sample bias circularly modulated structures 
in the topographic image have been interpreted 
as Friedel oscillations around the ionized donors induced by the accumulated 
electrons in the space-charge region \cite{Wielen96}. Furthermore, in $n$-doped InAs the 
scattering states of ionized dopants at low temperatures 
have been explored only recently \cite{Wittneven98}. 

In addition to the delocalized features in the 
subsurface region, localized features have also been found which were 
attributed to a local change of the 
electronic structure around dopants at the GaAs (110) surface \cite{Zheng94}. 
Remarkably, {\it ab initio} calculations predict that the additional
electron of the substitutional Si$_{\rm Ga}$ atom is 
trapped by a localized midgap level due to a local modification 
of the electronic structure around the Si atom \cite{Wang93}.
Furthermore, different kinds of dopant-induced features 
and surface defects have been distinguished by voltage 
dependent imaging of the occupied and unoccupied 
electronic states \cite{Domke96}.

Apart from studies of the segregation or adsorption of group III or 
group IV elements on the Si (111) surface,
most of the work on the local electronic structure around individual 
dopant atoms focussed on doped III-V semiconductors rather than on 
elemental semiconductors like Si, although these 
surfaces have been investigated 
in great detail by STM and scanning tunneling spectroscopy (STS)
\cite{Becker93}. The question of 
how the dopant atoms
are spatially distributed is of crucial interest for the investigation of
the metal-insulator transition in Si doped with phosphorus 
and/or boron \cite{HvL}.  
In a previous report  we have shown that individual 
P donors at the ($2\times 1$) reconstructed (111) 
surface of $n$-doped 
Si can be identified  by voltage dependent imaging and STS
 at room temperature \cite{Trapp97,Trapp99}.
Electronic surface states energetically located in the bulk band gap 
pin the Fermi level at $0.4$ eV above the 
top of the valence 
band at the surface. Consequently, the bands are bent upward towards the surface in 
contrast to GaAs (110) where the bands remain flat at zero bias. 
In Si:P the Coulomb 
potential around the ionized donor causes a local down-shift of 
the electronic density of states (DOS) and the donor appears as a protrusion 
at positive sample bias and as an indentation at negative bias.

In order to further investigate the change of the local electronic structure 
around dopant atoms at the Si (111) surface we performed 
STM and STS measurements on cleaved $p$-doped Si (Si:B) at room temperature.
Previous studies on the adsorption of B on Si (111) revealed 
different stages of B incorporation in the surface depending 
on coverage and thermal treatment \cite{Lyo,Bedrossian} in contrast 
to other group III adsorbates \cite{Becker88,Hamers}. 
For the cleaved Si:B (111) surface one might naively expect that 
apart from the change of the sign of the Coulomb potential the influence 
of the  negatively ionized B acceptor on the electronic structure can be explained 
in a way similar to Si:P. However, we will show that unlike in Si:P, the 
DOS in Si:B is strongly modified at the acceptor 
site possibly due to the different electronic configuration of 
substitutional B in Si compared to P.

\section{Experimental}

Measurements were performed with an Omicron STM in ultra-high vacuum (UHV)
at room temperature.
STM tips were prepared from electro-chemically etched tungsten wire and further 
cleaned in UHV by repeated cycles of annealing and consecutive Ar$^+$ sputtering.
Samples ($0.3 \times 4 \times 10$ mm$^3 $) with boron concentrations 
$N_A = 7 \cdot 10^{18} \rm{cm} ^{-3}$\ and $4.5 \cdot 10 ^{19} \rm{cm} ^{-3}$\ 
were cut from Czochralski-grown single crystals \cite{Wacker}
and cleaved {\it in situ}
to expose the (111) surface to the tip. STM images were acquired 
directly after cleavage without further heat treatment of the samples
to maintain the original dopant distribution  
at the surface.
Images were taken in the constant-current mode with the voltage
applied to the sample and the tip grounded. 
Hence, at positive voltages 
unoccupied electronic states of the sample 
are imaged whereas at negative voltages 
occupied states are imaged, implying that the DOS 
of the tip varies smoothly 
with energy. 
Scans of opposite polarity 
were acquired quasi-simultaneously by scanning each line forward 
and backward with reversed polarities. 

\section{Results and Discussion}
\subsection{Identification and Distribution of B Acceptors}

\begin{figure}
\centerline{\psfig{file=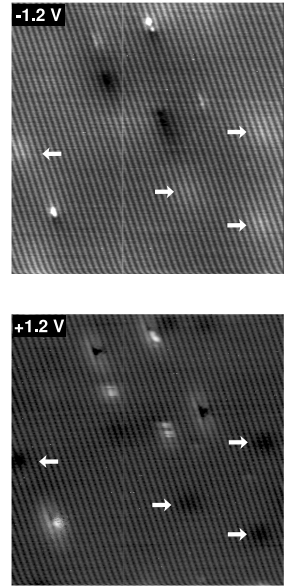,width=60mm}}
\vspace{4mm}
\caption[]{ STM images (30 nm $\times$ 30 nm) of the 
cleaved Si:B (111) surface 
($N_A = 7 \cdot 10^{18}$cm$^{-3}$) at negative 
and positive sample voltage $U$, 
tunneling current $I$ = 0.7 nA. White arrows indicate
B-induced features.} 
\label{fig1}
\end{figure}

\begin{figure}
\centerline{\psfig{file=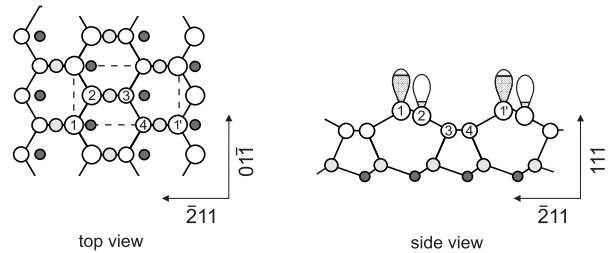,width=80mm}}
\vspace{4mm}
\caption[]{Schematic sketch of the  $2 \times 1$\  reconstructed (111) surface. 
The lobes indicate the occupied or unoccupied dangling- 
bond states at atom 1 or 2, respectively.}
\label{fig2}
\end{figure}

Fig. \ref{fig1} shows STM images 
of the cleaved Si(111) surface at +1.2 V and $-$1.2 V 
respectively. The bright rows are characteristic for the 
($2 \times 1$) reconstruction of the cleaved (111) surface 
as investigated in detail by Feenstra et al. \cite{Feenstra87}. 
The reconstruction is explained by a revised $\pi$-bonded 
chain model \cite{Pandey81,Badziag88} 
where the $p_z$ orbitals representing 
`dangling bonds' (DB) 
retain their local character and are therefore ideally suited for
imaging with an STM. Fig. \ref{fig2} shows a 
schematic sketch of this reconstruction where 
atoms marked 3 and 4 are fourfold coordinated and lowered, 
while the outer atoms 1 and 2 have one DB each which  
is mainly occupied at atom 1 and mainly unoccupied
at atom 2 \cite{Badziag88}.
In the STM only the elevated atoms 1 and 2 are imaged. 
At {\it negative} sample voltage electrons tunnel out of
the mainly occupied orbitals at position 1 which therefore appear as 
bright chains in the image, separated by dark stripes 
corresponding to atoms at positions 3 and 4. 
At low {\it positive} voltages the electrons are able to 
tunnel into the mainly unoccupied states located at atom 2 
which therefore appear as bright chains. The small shift of the rows
along the [\={2}11] direction is verified by comparing images of 
opposite polarity. 
At even higher positive voltages electrons
tunnel into unoccupied states that are located at the bonds between 
atoms 1 and 2 so that the rows
appear as a zigzag chain.
Several types of defects are observed. Those marked by arrows have a {\it characteristic}
voltage-dependent contrast. These defects appear as protrusions 
(bright) at $-$1.2 V in contrast to
images taken at +1.2 V where they appear as indentations (dark).
Hence, the contrast is exactly inverted with respect to Si:P.
These characteristic defects are observed for both 
investigated boron concentrations.
We count a total of 67 defects in a surface area of 21323 $\rm{nm} ^2 $\  for 
$N_A=7 \cdot 10^{18} {\rm cm}^{-3}$\ and  106 defects in 8357 $ \rm{nm} ^2 $\ 
for $N_A=4.5 \cdot 10 ^{19} {\rm cm}^{-3}$. The values correspond 
to dopant densities of 
$3.14 \cdot 10^{-3} \rm{nm} ^{-2}$\ and $1.26 \cdot 10 ^{-2} \rm{nm} ^{-2}$, 
in good agreement with the respective surface densities 
of $ 2.2 \cdot 10^{-3} \rm{nm} ^{-2}$\ and $1.4 \cdot 10 ^{-2} \rm{nm} ^{-2}$ 

obtained from the bulk concentrations,
where we assume that only atoms in the outermost layer 
(atoms 1 to 4, Fig. \ref{fig2}) give rise to a
contrast in the image. We therefore ascribe these features as 
being due to single B acceptors in the surface layer. 
 It is reassuring that structures with this specific 
voltage-dependent contrast
have not been found on the previously investigated
Si:P (111) surface \cite{Trapp97,Trapp99}.
Vice versa, the characteristic voltage-dependent contrast 
ascribed to P donors on Si:P
has not been found on the Si:B surface investigated here.
 
\begin{figure}
\centerline{\psfig{file=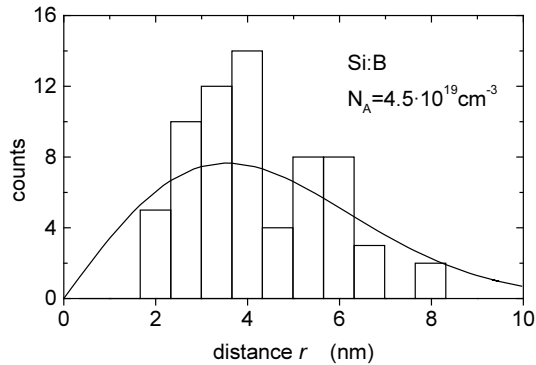,width=70mm}}
\vspace{4mm}
\caption[]{Distribution of B nearest neighbors as determined from 
STM images (histogram) and $f(r)$ derived for 
a random arrangement of B atoms (solid line).} 
\label{fig3}
\end{figure}

The identification of these defects as individual 
B dopants allows to check whether 
the dopant distribution in Si:B is random. 
The probability of finding the 
nearest neighbor B atom at a distance $r$\  from a given atom at the surface is  
$f(r)=2 \pi r \rho e^{-\rho \pi r^2}$, assuming a Poisson distribution
of dopants with the dopant surface-density $\rho$.
For the determination of the
nearest neighbor distances, B atoms being nearer to the image border
than to any other B atom have been discarded.
The distances $r$\ have been grouped into 0.665 nm wide 
intervals corresponding to the unit cell 
dimension along [\={2}11].
Fig. \ref{fig3} shows a histogram together with 
$f(r)$\ normalized to the area under the histogram. The behavior 
does not change considerably when the intervals are shifted by the row-to-row 
distance of 0.332 nm.  
We find reasonably good agreement between the statistical and experimental distribution, although
 there seems to be a cut-off at low $r$ similar to 
what has been observed for Si:P. 
The shortest distance between two 
B atoms was found to be $1.67 \: \rm nm$. This is in reasonable 
agreement with the distance
of $2.3\: \rm nm$\ that follows from the solubility limit 
$1.2\: {\rm at}\% \; (T<1400 \:{\rm K})$\ of B
in Si for a random distribution \cite{Oles}.
 The reduced counts at longer distances are likely to be due to the
restricted size of the STM images. For this reason, 
the statistics has been carried out for the higher concentration of
$N_A=4.5 \cdot 10^{19} {\rm cm}^{-3}$ only, because most of the images taken on 
the sample with $N_A=7 \cdot 10^{18}{\rm cm}^{-3}$
showed only one or two B defects. 
We conclude that B acceptors in Si:B are distributed randomly with a 
short-distance cut-off similar to P donors in Si:P.

\subsection{Electronic Structure of Single B Acceptors}

\begin{figure}
\centerline{\psfig{file=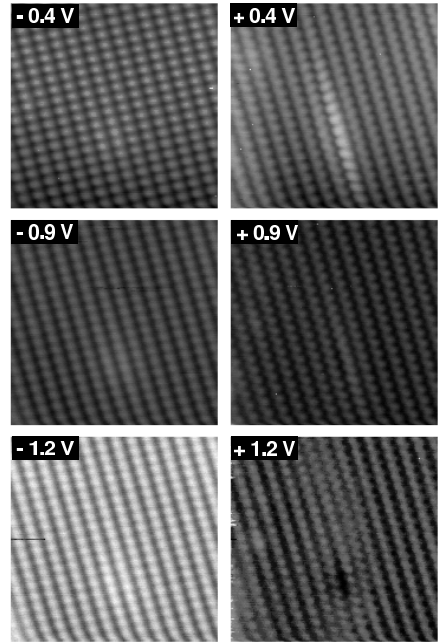,width=70mm}}
\vspace{4mm}
\caption[]{STM images ($7 \times 7$ nm$^2$) of the Si:B (111) surface 
($N_A = 7 \cdot 10^{18}$cm$^{-3}$) 
at different sample voltages, $I = 0.7$ nA.} 
\label{fig4}
\end{figure}

We now discuss the voltage-dependent contrast in more detail. 
Fig. \ref{fig4} shows STM images of a B-induced feature 
for different voltages.
At all negative voltages boron appears as an isotropic shallow protrusion 
with a diameter of $\approx$ 2 nm. 
In contrast, at positive 
voltages B appears as a needle-shaped protrusion at $+0.4$ V 
but can hardly be
distinguished from Si at $+0.9$ V. Further increase of the voltage 
leads to a reversal of the 
image contrast compared to $+ 0.4$ V and the acceptor appears as an 
indentation. This change in contrast with voltage 
is explained by a local change of the surface DOS around the acceptor. 

\begin{figure}
\centerline{\psfig{file=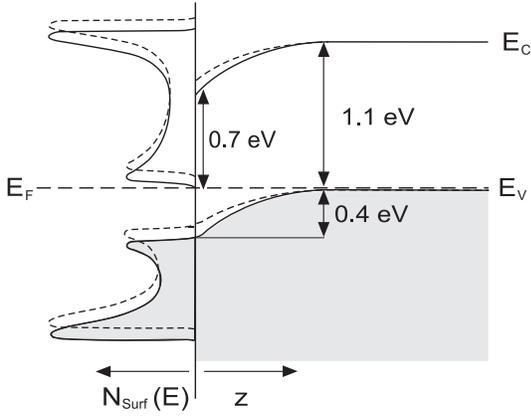,width=70mm}}
\vspace{4mm}
\caption[]{Surface density of states and energy bands for 
$p$-doped Si (111) at 
the Si position (solid line) and at the acceptor position (dashed line). 
Hatched areas indicate occupied electronic states of Si.} 
\label{fig5}
\end{figure}

Fig. \ref{fig5} shows the surface DOS for the $p$-type cleaved Si (111) surface 
together with the near-surface bulk energy bands as inferred from 
photoemission and STS experiments \cite{Feenstra87,Allen,Himpsel}.  
In $p$-type Si, the acceptor level at 45 meV above the valence 
band edge is occupied at room temperature and the boron atom
is negatively charged. This occurs for B acceptors at the surface as well, 
due to the low hole binding energy of 25 meV estimated for high carrier 
concentrations \cite{Brum84}. 
In addition, due to the presence of electronic surface states in the 
bulk band gap (Fig. \ref{fig5}, solid line), 
holes (majority carriers) are accumulated at the 
surface and the positive charge 
is compensated by the formation of a hole depletion layer of depth $d$. 
The negative space charge of the depletion layer 
gives rise to a downward band bending towards the surface.
Furthermore, the Fermi level is pinned 0.4 eV 
above the valence band edge at the surface 
almost independent of dopant concentration
for moderately doped Si \cite{Allen,Himpsel}.
For a boron concentration of 
$N_A = 4.5 \cdot 10^{19}$ cm$^{-3}$ 
and a band bending of $-$0.4 V a depletion-layer depth $d= 3.4$ nm 
and a surface charge corresponding to 
$1.5 \times 10^{13}\: e$cm$^{-2}$ are estimated \cite{Monch95}. This 
charge corresponds to a large hole density 
of $2.6 \times 10^{20} \: {\rm cm}^{-3}$ near the surface
owing to the fact that the surface states decay into the volume 
with a short decay length of 1 - 2 lattice constants \cite{Tersoff84}.

Adopting the simple description which worked successfully for Si:P,
the surface DOS at 
the location of the boron atom is energetically shifted {\it upward} 
due to the screened negative 
Coulomb potential of the B acceptor (Fig. \ref{fig5}, dashed line).
For the occupied states below $E_F$\ this upward shift leads to a local
increase of the surface DOS near the Fermi level. 
At negative voltages, electrons near the 
Fermi level of the sample
have the highest transmission factor for 
tunneling into unoccupied states of the tip.
Therefore, at the site of the B atom
 more electrons are able to tunnel
into the tip compared to Si sites. Thus, B appears as a
protrusion for {\it all} negative voltages.

At positive voltages, electrons at energies near the 
Fermi level of the tip
have the highest probability for tunneling into 
unoccupied surface states of the sample.
At $+0.4$ V there are more states available at the position 
of B which therefore appears as a bright protrusion.
 At $+0.9$ V the surface DOS of B and Si are alike so 
that B and Si can hardly be distinguished. 
At even higher voltages ($+ 1.2$ V)
there are more surface states on Si,
leading to a higher tunneling probability at the Si position
than at the B position and B appears as an 
indentation. Hence, in this simple picture of a locally shifted DOS the 
contrast changes from bright to dark with increasing positive voltage 
as experimentally observed. 

\begin{figure}
\centerline{\psfig{file=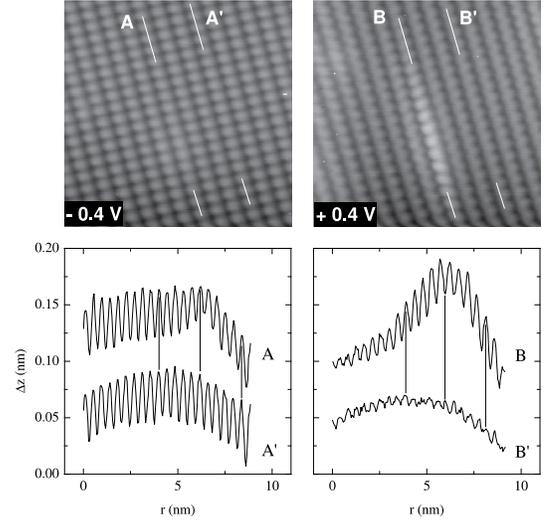,width=70mm}}
\vspace{4mm}
\caption[]{STM images (7 nm $\times$ 7 nm) of the boron-induced 
feature for $U = \pm 0.4$ V 
(cf. Fig. \ref{fig4}) and height profiles $\Delta z(r)$ 
along the indicated lines. 
In the scan along B one orbital is missing at the position of boron
compared to the scan along B'. } 
\label{fig6}
\end{figure}

A more detailed analysis of the image taken at $+0.4$ V shows that 
apparently one unoccupied
orbital in the row containing the defect is 
missing, compared to the 
image for $U = -0.4$ V (Fig. \ref{fig6}). 
The same behavior was found for 
$\pm 0.6$ V bias. The fact that the 
orbital is present for higher voltages 
confirms that the feature is not due to a surface vacancy.
The missing orbital is better seen in 
the cross-section lines where each maximum
corresponds to an unoccupied orbital (Fig. \ref{fig6}). 
At the position of the B acceptor 
a minimum appears  
at a position where actually a maximum is expected. 
Counting the number of maxima along
the two lines B, B', i.e. along the [01\=1] direction, 
one maximum is missing in the line taken across the defect (B) 
compared to a scan away from the acceptor (B').
We stress that this behavior was found for about 50 \% 
of all B-induced features while 50 \% appeared regular.

In Si:P the appearance of an additional orbital at the position 
of the P donor at U $= - 0.4$ V
\cite{Trapp97,Trapp99} 
was explained as being due to Si surface states just above $E_F$, 
which are usually unoccupied
but are shifted to below the Fermi level by the Coulomb potential of the 
{\it positively} charged P donor.
These states refer to the dangling-bond orbital at atom 2 
(cf. Fig. \ref{fig2}). 
Therefore, this orbital
becomes occupied and appears as an additional ``atom'' at negative voltages.
As is  apparent from Fig. 5, 
a corresponding explanation invoking simply a local
shift of the surface DOS
cannot be found to account for a missing orbital at the acceptor site
in Si:B. We believe that this is 
due to different substitutional
positions of boron at the surface.
Rather, a qualitative change of the surface DOS at the B site must be
invoked.
For instance, the trivalent B located at position 1 or 2 would 
participate in three covalent bonds although the preferred 
flat $sp^2$-hybrid
configuration cannot be realized. In contrast, B at position 
3 or 4 has to satisfy four bonds making a charge transfer from 
adjacent Si atoms very likely. Such an effect of the B substitutional 
site on the electronic structure has been reported for 
B deposited on Si (111) where B does indeed occupy immediate
subsurface positions \cite{Lyo,Bedrossian}.
In addition, the Si-B
bond length is about 12 \% shorter than the Si-Si bond length
mainly due to the smaller covalent radius of B 
compared to Si \cite{Bedrossian}.
This will lead to a relaxation of the surface structure 
around B which could be  connected with a
charge transfer between Si and B and/or a 
reformation of the unoccupied DB at/nearest to the B site 
towards a more back-bonded type. 
A missing orbital in the STM image at $+0.4$ V could either be 
explained by a complete occupation of the orbital at position 2, which then would 
not be imaged for $U > 0$, or by a reformation of the DB towards a more back-bonded character. 
\begin{figure}
\centerline{\psfig{file=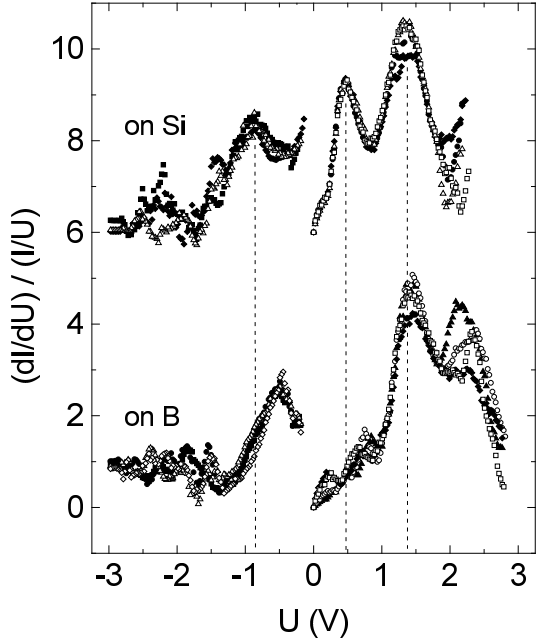,width=70mm}}
\vspace{4mm}
\caption[]{Normalized tunneling spectra taken away from and on top of 
the B-induced feature, respectively. Different symbols indicate 
different tunneling spectra calculated from an averaged set 
of several $I(U)$ curves.} 
\label{fig7}
\end{figure}

Because only atoms 1 or 2 are imaged 
the effect of B at position 3 or 4 on the image contrast is 
presumably weaker 
because the acceptor at position 3 or 4 can only be ``imaged'' via an electronic
interaction to Si atoms at positions 1 or 2.
Hence, we conclude that the effect of a missing orbital 
is presumably due to B located at positions 1 or 2. This is in  
good agreement with the fact that we found a missing 
orbital at roughly 50 \% of the B sites.
We emphasize
that the topological structure of the B-induced feature 
clearly requires theoretical calculations. Moreover, a more 
sophisticated model has to include many-body effects which are 
important for highly localized electronic states \cite{Hamers}.

The local change of the DOS at the acceptor site
is confirmed by scanning tunneling spectroscopy (STS). 
During $I(U)$ data acquisition the tip was retracted from 
the surface by $0.05\: \rm{nm/V}$\ to compensate for 
the exponential increase of current with voltage.
Se\-ve\-ral $I(U)$ curves were averaged to reduce the scatter of the data.
As demonstrated earlier,  
the logarithmic derivative $(dI/dU)/(I/U)$\ is independent 
of the distance and is related to the 
surface DOS with $U=0$  corresponding to $E_{\rm F} $  \cite{Feenstra87} .
Due to the numerical evaluation of $(dI/dU)/(I/U)$\ 
data points around $U=0$\ were omitted because 
division by zero leads
to unreasonable results, while $(dI/dU)/(I/U)|_{U=0} = 1$.
Fig. \ref{fig7} shows spectra taken 
above a Si atom away from the acceptor which exhibits 
peaks at $U = -0.9, +0.5 , +1.4 \: \rm V$\ in agreement with  
previous results for $p$-doped Si where similar peaks were 
observed at somewhat 
lower voltages $ U = -1.0, -0.35, +0.17, +1.25$ V 
due to the lower 
doping level \cite{Feenstra87} compared to the present samples. 
These peaks can be attributed 
to the surface DOS of Si observed in photoemission experiments 
\cite{Allen,Himpsel}. A peak expected
at $U = - 0.35\: \rm V$\ corresponding to the occupied DB 
states could not be resolved due to the numerical difficulties 
around $U = 0$. 

In comparison, the spectrum taken above the acceptor 
is strongly modified, apart from the peak at $U = +1.45\: \rm V$ 
which is slightly shifted by $\Delta U \leq 0.1$ V compared to the 
peak on Si. The most striking result is the strong 
reduction of the large peak 
observed on Si at $+0.5$ V. The corresponding unoccupied 
electronic states are located at
the DB orbitals at position 2. Strong modifications 
of the STS spectra have also been reported for B deposited on 
Si (111) \cite{Lyo,Bedrossian}. In the present case the reduction 
of the surface DOS above $E_{\rm F}$ confirms the
absence of a localized DB state at the B site and the 
effect of a missing orbital in the STM image at $U = 0.4 - 0.6 \: \rm V$.
Furthermore, the peak at $-0.9$ V representing the bottom of the 
occupied surface band is considerably shifted upward by $\approx 0.5$ 
V at the B site. 
Since at negative bias, states near $E_{\rm F}$ 
of the sample dominate the tunneling current this confirms the 
bright image contrast at the B site observed for all negative 
voltages.       

A further question concerns the apparent anisotropic shape of the B defect.
As mentioned above, at $- 0.4$ V the extension of the B induced 
feature seems to be isotropic while it appears strongly elongated
along the [01\=1] direction at positive voltage. Such an effect could be due 
to a different electronic interaction between Si and 
B along the $\pi$-bonded chains compared to the perpendicular direction 
and needs further investigation.
We mention that the P-induced feature on Si:P appeared to be 
isotropic, although its overall extension was smaller.
An explanation for the larger extent
of the B-derived features could be a different distance between the tip and the surface 
compared to Si:P caused by a higher tunneling current of 0.7 nA compared 
to 0.3 nA in the latter case.

\section{Summary}

The Si (111) ($2\times 1$) surface of cleaved B-doped single crystals has been 
investigated by STM. The individual B acceptors have
been identified by a voltage-dependent contrast 
and have been further characterized
by STS. Similarly to Si:P the general change in contrast is 
governed by the local shift of
the surface DOS due to the Coulomb potential of the charged 
dopant atom. In addition, the local change of the atomic and 
electronic structure around the B acceptor 
gives rise to a missing orbital observed in 50 \% of the 
images taken at $U= +0.4$ V. These acceptors are presumably 
located at the upper positions 1 and 2 of the $(2 \times 1)$ 
reconstructed surface.
The unoccupied DOS of B at the surface  as determined by STS
is strongly 
reduced around 0.4 eV, in contrast to Si:P where such 
drastic changes are not observed. This shows that the electronic 
configuration of substitutional dopants 
has a decisive influence on the local electronic structure.   

The fact that in Si:B and Si:P the surface-densities of dopants estimated
from the STM data are in good agreement with the surface densities
derived from the bulk concentrations strongly  confirms the assumption 
that only dopants in the outermost layer are imaged. This is in strong contrast
to GaAs, where features arising from dopants in several subsurface layers
are observed \cite{Zheng94,Johnson93}. Presumably, the screening of dopants 
is very different in doped Si and needs to be investigated further.
\acknowledgements

This work was supported by the Deutsche Forschungsgemeinschaft
through the Graduiertenkolleg "Kollektive 
Ph\"anomene im Festk\"orper"and the Sonderforschungsbereich 195. 
We would like to thank T. Trappmann and 
M. Wenderoth for useful discussions.

\end{document}